\documentclass[twocolumn,aps,prd,amsmath,amssymb,nofootinbib,longbibliography]{revtex4-1}
\pdfoutput=1
\usepackage[english]{babel}
\usepackage{amsmath,amssymb,amsbsy,amstext,amsthm,booktabs,simplewick,exscale,relsize,slashed,graphicx,amsfonts,upgreek}
\usepackage{tabu,multirow,color,dcolumn,bm,enumerate}
\usepackage{tikz}
\usetikzlibrary{decorations.pathmorphing,,decorations.markings}

\def\mpl{m_{\rm{pl}}}

\begin{document}


\title{Higgs portals to pulsar collapse}

\author{Joseph Bramante}
\address{Department of Physics, University of Notre Dame, IN, USA}

\author{Fatemah Elahi}
\address{Department of Physics, University of Notre Dame, IN, USA}

\begin{abstract}
Pulsars apparently missing from the galactic center could have been destroyed by asymmetric fermionic dark matter ($m_X = 1-100$ GeV) coupled to a light scalar ($m_{\phi}= 5-20$ MeV), which mixes with the Higgs boson. We point out that this pulsar-collapsing dark sector can resolve the core-cusp problem and will either be excluded or discovered by upcoming direct detection experiments. Another implication is a maximum pulsar age curve that increases with distance from the galactic center, with a normalization that depends on the couplings and masses of dark sector particles. Finally, we use old pulsars outside the galactic center to place bounds on asymmetric Higgs portal models.
\end{abstract}

\maketitle

\setcounter{footnote}{0}

\section{Introduction}
\label{sec:intro}

 
Although there is compelling evidence for the existence of particulate dark matter (DM) from astronomical observations, the precise nature of DM remains elusive. One possibility is that a weakly interacting massive particle (WIMP) froze out during the primordial epoch with a picobarn annihilation cross-section. However, the WIMP paradigm does not explain why the relic abundance of DM is comparable to that of baryonic matter: $\Omega_{\rm DM}  \sim 5  \Omega _{\rm  B}$ \cite{Hinshaw:2012aka,Ade:2013zuv}. The baryonic relic abundance itself was determined by a baryon asymmetry arising at some temperature hotter than the baryon-anti-baryon freeze-out temperature in the early universe, $ T \gtrsim 10 $ MeV. While the exact source of the baryon asymmetry is unknown, the Sakharov conditions stipulate that particle charge (C) and charge-parity (CP) must be violated out of thermal equilibrium to generate a particle asymmetry. There are numerous models that meet these conditions for baryons, including out of equilibrium decays \cite{Weinberg:1979bt,Fukugita:1986hr} and the Affleck-Dine mechanism \cite{Dine:1995kz,Riotto:1999yt,Enqvist:2003gh,Dine:2003ax,Allahverdi:2012ju}. 

Because the source of C and CP violation responsible for the baryon asymmetry is unknown, it is plausible, and indeed the coincidence $\Omega_{\rm DM}  \sim 5  \Omega _{\rm  B}$ suggests, that the dark sector participated in the generation of the baryon asymmetry. In the simplest dark asymmetry models (see e.g. \cite{Barr:1990ca,Kaplan:1991ah,Farrar:2005zd}), DM is charged under a quantum number that is composed of or identical to baryon number, and the DM abundance is set by the same process that sets the baryon abundance. Such models usually imply a DM mass range $m_X = 1-15 ~\rm{GeV}$ and collectively fall under the rubrik of asymmetric dark matter (ADM) (see Refs.~\cite{Kaplan:2009ag,Petraki:2013wwa,Zurek:2013wia} for recent reviews). 

After the dark asymmetry is generated, the final ADM relic abundance will depend on having a large annihilation cross-section at freeze-out. ADM freeze-out cross-sections must be larger than that of WIMP dark matter, because if the dark asymmetry provides for most of the dark matter relic abundance, ADM annihilation cross-sections must exceed the standard picobarn-size DM freeze-out cross-section. Otherwise, ADM would freeze-out to an overabundance and collapse the universe. Altogether, these relic abundance requirements motivate ADM coupled to light mediators, because a picobarn cross-section is difficult to attain using heavy ($\gtrsim 100$ GeV) mediators, while remaining consistent with collider bounds \cite{ATLAS:2012ky,Khachatryan:2014rra}\footnote{DM-nucleon couplings for $m_X=0.01-1$ GeV and sub-TeV mediators are most constrained by quarkonium$\rightarrow$invisible decays at Belle and BES \cite{Fernandez:2014eja}.} on low mass dark matter (i.e. $m_X = 1-15 ~\rm{GeV}$). However, collider bounds on light DM can be evaded if DM annihilates to SM particles through a light mediator mixed with a SM boson \cite{Lin:2011gj}.
 
In addition to evading collider bounds and matching cosmological requirements, light mediator dark matter also fits a number of astrophysical anomalies \cite{Hooper:2008im,ArkaniHamed:2008qn,Pospelov:2008jd,Kaplan:2009ag,Buckley:2009in,Loeb:2010gj,Tulin:2012wi,Tulin:2013teo}, including the core-cusp problem -- some dwarf galaxies exhibit unexpectedly cored DM density profiles \cite{Oh:2010ea,KuziodeNaray:2007qi} -- and the too big to fail problem -- there is a dearth of massive Milky Way satellite galaxies \cite{Sawala:2010zw,BoylanKolchin:2011de,BoylanKolchin:2011dk}. 

More recently, radio surveys of the galactic center have discovered a missing pulsar problem \cite{Dexter:2013xga,Chennamangalam:2013zja}. Based on the number of progenitor stars which collapse into pulsars, around $10$ young pulsars were expected to have already been found within the Milky Way's central parsec. However, so far no young pulsars have been observed. 

In this paper, we demonstrate that the missing pulsar problem specifically motivates asymmetric fermionic dark matter coupled to the SM through a light Higgs portal mediator, which would collect in and collapse pulsars at the galactic center, yet remain consistent with older pulsars seen in less DM-dense regions. The structure of this paper is as follows: in the remainder of the introduction, we provide a review of the missing pulsar problem. In section \ref{sec:model}, we introduce the Higgs portal model, and catalogue constraints on it from relic abundance, primordial nucleosynthesis, direct detection, and Higgs invisible width measurements. In section \ref{sec:collapse} we determine what Higgs portal parameter space could simultaneously explain the absence of pulsars at the galactic center, while remaining consistent with old pulsars observed near the solar position. Section \ref{sec:agecurves} gives the expected maximum age of pulsars as a function of distance from the galactic center. We present concluding remarks in section \ref{sec:conclusions}. Appendix \ref{app:pulsars} details estimates of the expected pulsar population at the galactic center. Appendix \ref{app:snx} addresses the velocity dependence of the DM-nucleon cross-section.

\subsection{The missing pulsar problem}
\label{sec:missingpulsars}

Radio surveys of the galactic center have not found as many pulsars as expected \cite{Dexter:2013xga}. Pulsars are rapidly rotating, strongly magnetized neutron stars with spin period $P$ and approximate spin lifetime $\tau_{p} \lesssim  P/2\dot{P}$, formed in the supernova collapse of (8-20 solar mass) heavy progenitor stars (see \cite{Tauris:2015bra} for a review). There are about $300$ heavy progenitor stars within a parsec of the galactic center \cite{Bartko:2009qn}, which signify the presence of at least $500$ young (0.01-100 Myr) pulsars in the same region \cite{Dexter:2013xga,Chennamangalam:2013zja,Pfahl:2003tf}.

In addition to these young pulsars, one thousand old ($\sim$ Gyr) millisecond pulsars (MSPs) are expected to reside in the central parsec. Millisecond pulsars form when a neutron star in a binary system accretes material and angular momentum from its binary companion, thereby spinning up to millisecond pulse periods. The dense stellar environment of the galactic center should host many binary systems; extrapolating from the population of millisecond pulsars observed in star-dense globular clusters (e.g. Terzan5 has $\sim 100$ MSPs \cite{Bagchi:2011hs}), an estimated $500-10^4$ millisecond pulsars should abide in the central few parsecs \cite{FaucherGiguere:2010bq}.

In addition to young and millisecond pulsars, there are also magnetars, which have a $\gtrsim \! 100$ times stronger magnetic field and slower pulses than most pulsars. To date, a single magnetar 0.1 parsecs from the galactic center, is the only pulsar discovered inside the central 50 parsecs \cite{Mori:2013yda,Kennea:2013dfa}. Prior to the measurement of this magnetar's radio pulse dispersion \cite{Spitler:2013uva}, it was reasonable to assume that pulsars had not been seen in the galactic center, because their radio pulses were smeared out by compton scattering off a dense intervening column of electrons. 

However, using radio pulses from the magnetar, the density of electrons between the GC magnetar and earth has been measured. This measurement implies that, with the most conservative assumptions, prior radio surveys of the galactic center \cite{Macquart:2010vf,Wharton:2011dv} should have found at least 10 young pulsars and 4 millisecond pulsars \cite{Dexter:2013xga}. (Note that only about a tenth of existing pulsars will have pulses that beam towards earth.) A Bayesian analysis of the missing young pulsar problem found that the 6.6 GhZ galactic center survey \cite{Bates:2010wb} puts a $99 \%$ upper limit of 200 young pulsars in the central parsec, comparable to 500 expected using high mass progenitor data \cite{Chennamangalam:2013zja}. In  Appendix \ref{app:pulsars} we provide more detail on methodologies for estimating the galactic center pulsar population. There are a number of possible explanations for the missing pulsar problem -- the $300$ heavy progenitor stars observed inside the central parsec could be a recent anomaly, and not indicative of heavy progenitor populations 100 Myr ago \cite{Dexter:2013xga}. Some estimates of the galactic center millisecond pulsar and young pulsar populations assume a typical stellar initial mass function in the galactic center (see Appendix \ref{app:pulsars}). The actual initial mass function of the galactic center may be ``top heavy" \cite{Nayakshin:2005re}, resulting in more black holes and fewer pulsars produced through core collapse of GC high mass progenitor stars. In this paper, we focus on a different, intriguing possibility: asymmetric fermionic dark matter coupled to SM particles through a Higgs portal may collapse pulsars in the galactic center.

\section{Higgs portal mediators for pulsar collapse}
\label{sec:model}

While a number of astrophysical anomalies motivate fermionic dark matter coupled to an MeV-scale vector or scalar mediator \cite{Pospelov:2008jd,Lin:2011gj,Kaplinghat:2013yxa}, in the case of pulsar-collapsing dark matter (PCDM), scalar mediators and asymmetric DM are required. The dark matter must be asymmetric, so that it can collect in pulsars without annihilating to SM particles \cite{Bramante:2013hn,Bell:2013xk}. If it is coupled to a light scalar mediator, it will only have attractive self-interactions\footnote{Note that DM-mediator couplings of the form $g_D \bar X \gamma^\mu X \phi_\mu$ will be repulsive for $XX \rightarrow XX$ and $\bar{X}\bar{X} \rightarrow \bar{X}\bar{X}$ scattering, whereas scalar couplings $g_D \bar X X \phi$ produce purely attractive self-interactions. In both these cases the potential is Yukawa $|V| = g_D^2 \rm{exp}(-m_\phi r)/4 \pi r$. See \cite{Bellazzini:2013foa} for a treatment of pseudoscalar and axial couplings, which we do not consider here.}, permitting DM self-attractive forces to overcome Fermi degeneracy pressure and initiate black hole collapse at the center of pulsars.

Assuming the dark mediator $\phi$ is a real scalar, and $X$ is a dirac fermion, we add the following terms to the SM Lagrangian,
\begin{align}
\mathcal{L} &= \mathcal{L}_{\rm SM}  + i \bar X \displaystyle{\not} \partial X + \frac{1}{2} (\partial \phi)^2 - a \phi (|H|^2 - v^2/2) \nonumber
\\ & - b \phi^2 (|H|^2 -v^2/2)- g_D \phi \bar X X -m_\phi^2 \phi^2 - m_X \bar X X 
\label{eq:lagrang}
\end{align}
Thus $\phi$ and $X$ interact with the visible sector through $H-\phi$ mixing \cite{Foot:1991bp,Schabinger:2005ei,Patt:2006fw}. Note that we have taken $H \rightarrow h + v/\sqrt{2}$ as the Higgs vacuum expectation value (vev) after electroweak symmetry breaking and have shifted the $H-\phi$ coupling so that $\phi$ does not get a vev. In the low momentum limit, and assuming $ a, m_\phi \ll v, m_h$,  the effective coupling of $\phi$ to standard model particles is 
\begin{align}
\epsilon_h \equiv \sqrt{2} a v/ m_h^2 .
\end{align} 
 
The coupling of the light mediator to DM, $g_D$ (or equivalently $\alpha_D = g_D^2/4 \pi$), is bounded from below by relic abundance and CMB constraints. For thermal dark matter, relic abundance requirements put a lower bound on the annihilation cross-section of dark matter $ \langle \sigma v\rangle \gtrsim 10^{-25} \rm cm^3/\rm s$, so that the universe does not collapse. If the scalar mediator is lighter than dark matter, this implies
\begin{align}
\alpha_D \gtrsim 11 \times 10^{-5} \left(\frac{\langle \sigma v\rangle}{10^{-25} \rm cm^3 /\rm s} \right)^{1/2} \left(\frac{m_X}{\rm GeV}\right) \left(\frac{x_f}{20}\right)^{1/2}
\end{align}
where $m_X$ is the mass of DM, $x_f \equiv m_x/T_{\rm FO} $, and $T_{\rm FO}$ is the freeze-out temperature of DM \cite{Lin:2011gj}. For asymmetric dark matter, requiring that $\Omega_x h^2 > 0.11$ (since the relic abundance is assumed to be generated by a particle asymmetry), the annihilation cross-section should be  $\langle \sigma v \rangle > 4.5 \times 10^{-26} \rm cm^3/s$ for Dirac fermions \cite{Steigman:2012nb}.

\begin{figure}[h!]
\begin{center}
\includegraphics[scale=1.12]{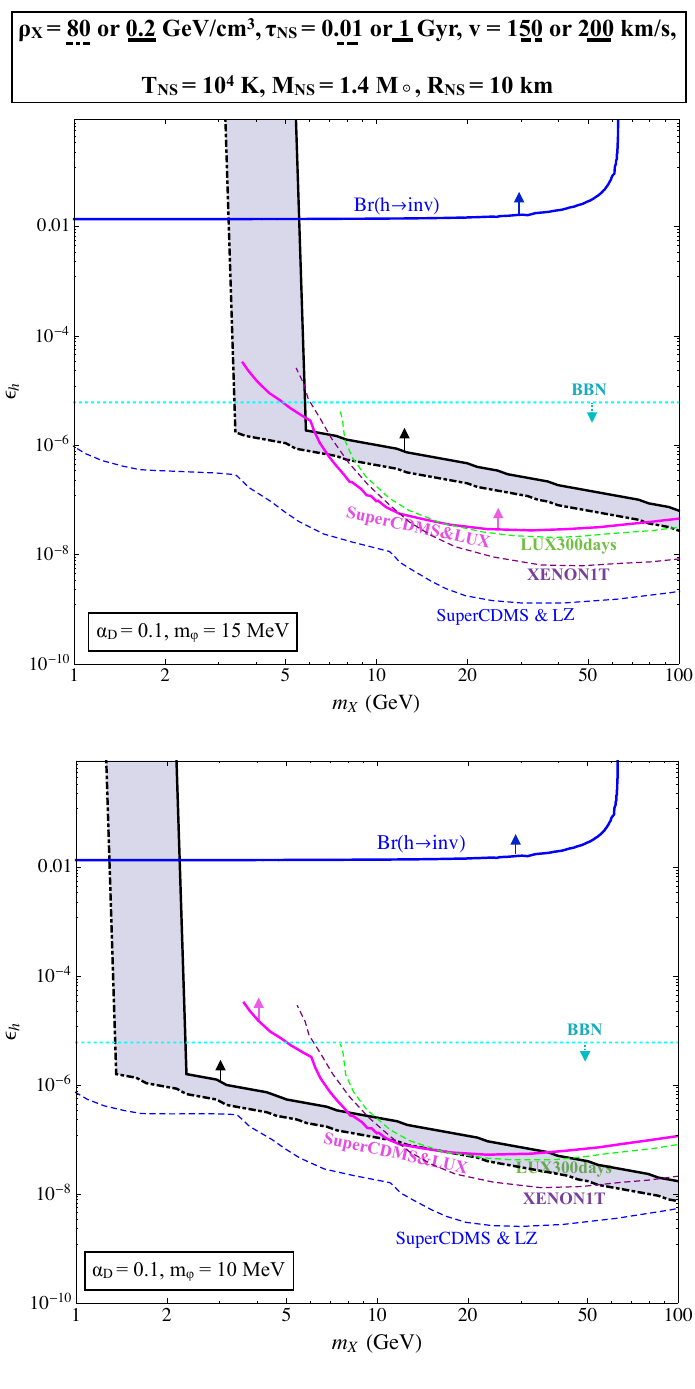}
\end{center}
\caption{Direct detection, cosmological, and collider bounds on Higgs portal dark matter are shown as a function of $\epsilon_h \equiv \sqrt{2}av/m_h^2$, along with viable galactic center pulsar-collapsing parameter space, shaded grey. Space above the solid black, blue and pink lines are excluded by Gyr old pulsars outside the GC, constraints on the Higgs invisible decay width, and direct detection, respectively. Constraints from BBN exclude space below the dotted cyan line, but these can be removed by adding decay modes for $\phi$, Ref.~\cite{Kouvaris:2014uoa}. The DM density ($\rho_X$), pulsar lifetime ($\tau_{NS}$), velocity dispersion ($v$), and pulsar temperature ($T_{NS}$), mass ($M_{NS}$), and radius ($R_{NS}$) used to plot the solid and dotted-dashed black lines are given at the top of the plot. Note that parameter space along the dotted-dashed (solid) black line would collapse pulsars older than $\sim10^7$  ($\sim10^6$) years inside the Milky Way's central 500 parsecs, while permitting up to $\sim10^{10}$ ($\sim10^9$) year old pulsars outside of the galactic center.}
\label{fig:epsvsmx}
\end{figure}

Because the mediator $\phi$ must decay before big bang nucleosynthesis, there is a lower bound on its coupling to SM particles. Big bang nucleosynthesis (BBN) takes place in the CMB radiation era after about a second. In this era, any extra particles that can decay and inject energy lead to an enhancement in the nuclear interaction rate. In particular, a light mediator with a mass of $1-100 ~\rm MeV$ is below the $^4\rm He$ binding energy; therefore it would disassociate deuterium and lithium isotopes, while leaving $^4\rm He$ relatively undisturbed. Thus there is a constraint on any light degree of freedom besides those of the SM \cite{Jedamzik:2009uy}. To avoid these constraints, we require that $\phi$ decay before BBN. For $m_\phi < 2m_\mu$, the total width of $\phi$ is 
\begin{align}
\Gamma_\phi & =  \frac{ \epsilon_h^2  m_e^2 m_\phi}{ 16 \pi v^2 } \left(1- \frac{4m_e^2}{m_\phi^2}\right)^{3/2}  \Theta( m_\phi - 2m_e)  \nonumber
\\ & +  \frac{\epsilon_h^2 \alpha_D^2 m_\phi^3}{512 \pi v^2} \left[ A_1 + \left(\frac{8}{3}\right) A_{1/2} + \left(\frac{1}{3}\right) A_{1/2} \right]^2,
\end{align}
where the second contribution is for $\phi \rightarrow \gamma \gamma$, with $A_1= -7$ from the $W$ loop, and $A_{1/2} = 4/3$ is the contribution from heavy quark loops \cite{Wise:2014jva}. In Figure \ref{fig:epsvsmx}, the dark cyan dotted line indicates parameter space where the lifetime of $\phi$ is less than a second. Note that even this constraint can be lifted entirely by coupling $\phi$ to additional light dark states (e.g. sterile neutrinos \cite{Kouvaris:2014uoa}).

$\phi-H$ mixing also contributes to the invisible Higgs decay width. The additional contributions to the Higgs invisible decay width from \eqref{eq:lagrang} are 
\begin{align}
\Gamma_{h_{\rm inv}}  &= \frac{\alpha_D  \epsilon_h^2 m_h}{2} \left(1- \frac{4 m_X^2}{m_h^2} \right)^{3/2}  +\frac{ b^2 v^2}{8 \pi m_h} \left(1 - \frac{4 m_\phi^2}{m_h^2} \right)^{1/2}.
\end{align}
Requiring the Higgs$\rightarrow$invisible branching ratio be less than $ 40\%$ (from e.g. \cite{Zhou:2014dba}) of the total SM Higgs width ($\Gamma_H \sim 4.1 ~ \rm MeV$) puts an upper bound on $\epsilon_h$. Since we only consider $m_\phi \ll m_h$, the constraint on the higgs invisible decay requires $b \ll 0.013$. Assuming the contribution from $b$ is negligible, in Figure \ref{fig:epsvsmx} we indicate the constraint on $\epsilon_h$ from restricting the Higgs invisible width to be less than $40\%$ of the SM expectation \cite{Aad:2013oja,Aad:2014iia,Chatrchyan:2014tja}.

DM-nucleon scattering experiments such as LUX \cite{Akerib:2013tjd}, XENON100 \cite{Aprile:2013doa}, and SuperCDMS \cite{Agnese:2014aze} also constrain Higgs portal parameter space. A calculation of DM-nucleon scattering is given in Appendix \ref{app:snx}, in which we include the full dependence of the scattering amplitude on $m_\phi$ and the velocity of DM in the nucleon's rest frame. Figure \ref{fig:epsvsmx} shows the current bounds \cite{Akerib:2013tjd,Agnese:2014aze} (solid magenta line) and future reach \cite{Cushman:2013zza} (dashed lines) of DM direct detection experiments.

\section{Collection and collapse of Higgs portal dark matter in pulsars}
\label{sec:collapse}

Another possible method for detecting asymmetric Higgs portal dark matter, is to find pulsars imploding (or having imploded) in regions where ambient dark matter is dense and can rapidly collect in pulsars. After enough asymmetric dark matter has been captured and thermalized in a pulsar, it may form a black hole that swallows the pulsar \cite{Goldman:1989nd,Kouvaris:2007ay,Casanellas:2009dp,McCullough:2010ai,deLavallaz:2010wp,Kouvaris:2010vv,Kouvaris:2010jy,McDermott:2011jp,Kouvaris:2011fi,Kouvaris:2011gb,Casanellas:2011qh,Lopes:2011rx,Casanellas:2012jp,Guver:2012ba,Bramante:2013hn,Bell:2013xk,Goldman:2013qla,Jamison:2013yya,Bertoni:2013bsa,Bramante:2013nma,Kouvaris:2013kra,Bramante:2014zca,Zheng:2014fya}. Usually, fermions will not form a black hole until $N_{ch} \sim (\mpl /m_X)^3$ particles have agglomerated, where $N_{ch}$ is the Chandrasekhar limit for fermionic matter. The asymmetric Higgs portal model we consider has attractive self-interactions (from $g_D \phi \bar X X $) that counteract Fermi degeneracy pressure, effectively decreasing $N_{ch}$. Indeed, attractive self-interactions are required for fermionic DM-induced pulsar collapse, because the maximum number of DM particles captured by a neutron star in a gigayear, $\sim 10^{41} (\rm{GeV}/m_X)$, is much smaller than the Chandrasekhar number, $\sim 10^{57} (\rm{GeV} /m_X)^3$.

\subsection{DM collection}

Dark matter particles are captured in neutron stars at a rate given by \cite{Goldman:1989nd,Kouvaris:2007ay,Bell:2013xk,Bramante:2013nma},
\begin{align}
C_X = \sqrt{\frac{6}{\pi}} \left( \frac{\rho_X}{\bar{v}_X} \right) \frac{N_B \xi v_{esc}^2}{m_X} \left[1 - \frac{1-\rm{exp} (-B^2)}{B^2} \right] f(\sigma_{nX}),
\end{align}
where $\rho_X$ is the DM density around the neutron star, $\bar{v}_X \sim 200 ~\rm{km/s}$ is the velocity dispersion of the dark matter-neutron star system, $N_B = 1.2 \times 10^{57}$ is the number of nucleons in a $1.4 M_\odot$ neutron star, $v_{esc} \simeq 0.7$ is the escape velocity from the surface of a neutron star, and $\xi = ~\rm{Min}[m_X/(0.2~\rm{GeV}),1]$ is a factor accounting for Pauli blocking of DM-nucleon scattering. Because the dark matter will have a semi-relativistic momentum when falling through the neutron star's rest frame, the size of its momentum $p_X \sim m_X$ indicates that heavy ($m_X \gg m_B$) dark matter may not transfer enough momentum to be captured by the neutron star. The term in square brackets, with the quantity $B^2 = 6 v_{esc}^2 m_X m_B / \bar{v}_X^2 (m_X + m_B)^2$, accounts for this diminution in capture of heavier dark matter. Finally, the term dependent on the dark matter-nucleon scattering cross-section, $\sigma_{nX}$, is given by $f(\sigma_{nX}) = \sigma_{sat} (1-\rm{exp}(-\sigma_{nX}/\sigma_{sat}))$. When the dark matter-nucleon scattering cross-section is small, $\sigma_{nX} \lesssim 10^{-45} ~\rm{cm^2}$, this function returns $\sigma_{nX}$, but as $\sigma_{nX}$ gets larger, the geometric cross-section per nucleon in the neutron star saturates. Note that for lighter dark matter, this saturation cross-section will depend on Pauli blocking \cite{Bell:2013xk,Bramante:2013nma}, $\sigma_{sat} = R_{NS}^2 / 0.45 N_B \xi$, where $R_{NS} \sim 10 ~\rm{km}$ is the neutron star radius. We detail our calculation of the dark matter-nucleon scattering cross-section $\sigma_{nX}$ for a given set of Higgs portal parameters ($\epsilon_h, m_X, m_\phi, \alpha_D$) in Appendix \ref{app:snx}. 

\subsection{DM state at collapse}

To determine the critical number of dark  fermions that will initiate black hole collapse in the neutron star interior, we use the Virial equation for a test particle at the edge of a sphere of fermions thermalized at the center of the neutron star,
\begin{align}
-2E_k + \frac{(\frac{4}{3} \pi)^{1/3} G \rho_B N_X^{2/3} m_X y^2}{m_\phi^2} + V_{Yuk} = 0,
\label{eq:virial}
\end{align}
where the first term is the virialized kinetic energy, the second term is the virialized gravitational potential from baryons in the neutron star, and the third term is the virialized form of the Yukawa potential. It is reasonable to assume all DM particles are thermalized, because results from \cite{Kouvaris:2012dz,Bertoni:2013bsa} indicate that for the parameter space under consideration, dark matter thermalizes inside the neutron star on timescales $\lesssim 10^5 ~ \rm{yrs}$. $N_X$ is the number of dark matter particles collected, and $V_{Yuk} = \sum_{r_j}  \alpha_D e^{-m_\phi r_j}(1/r_j + m_\phi) $, with $r_j$ being the inter-particle distance. Once $V_{Yuk} \gg 2E_k$, the dark matter will collapse into a black hole. In these calculations, we can neglect the dark matter self-gravity as being small (in what follows, see \cite{Bramante:2013nma} for more details). In Eq.~\eqref{eq:virial}, we have defined $y \equiv 1.6 m_\phi r / N_X^{1/3}$, which is the exponent of the Yukawa potential if only the nearest fermions to the test particle contribute to its potential; note that $x \equiv 1.6 r / N_X^{1/3}$ is the nearest-neighbor inter-fermion distance for $N_X$ fermions evenly distributed in a sphere of radius $r$. 

\emph{Strongly-screened.} If $y \gg 2$, only the nearest-neighbor particles will matter in determining the Yukawa potential. We will call this the ``strongly-screened" limit, the limit in which all but the nearest-neighbor fermions can be neglected. The strongly-screened virialized Yukawa potential is
\begin{align}
V_{Yuk}^{strong}=8\alpha_D \left( m_\phi e^{-y}/y + m_\phi e^{-y} \right),
\end{align}
where we assume eight nearest neighbor particles.

\emph{Coulombic.} If $y \ll 2$, on the other hand, the exponential piece of the Yukawa potential approaches unity, and the Yukawa potential becomes Coulombic. In other words, the Yukawa potential ($\propto e^{-m_\phi r_j}$) will not be suppressed inside radius $1/m_\phi$. The number density of dark matter fermions at the center of the star is given by $1/x^3$. Hence, the number of fermions contributing to the Coulomb-like potential inside radius $1/m_\phi$, is $N_{co} = 4 \pi / 3m_\phi^3 x^3$, which gives a virialized potential term,
\begin{align}
V_{Yuk}^{Coul}= 3 \alpha_D N_{co} m_\phi = 4 \pi \alpha_D m_\phi /y^3.
\end{align}

\emph{Degenerate.} Before the onset of collapse, the dark matter fermions collected inside radius $r$ will be degenerate if more than
\begin{align}
N_{deg} = 5 \times 10^{27} (r/\rm{cm})^3 (m_X /\rm{GeV})^{3/2}
\end{align}  
have collected, assuming an ambient temperature of $T=10^4 ~\rm{K}$. In this case the kinetic energy is given by\begin{align}
E_k^{deg} = (9\pi N_X/4)^{2/3} /2 m_X r^2 = (3 \pi^2)^{2/3} m_\phi^2 / 2 m_X y^2
\end{align}
and the pre-collapse radius of the dark matter, determined by solving Eq.~\eqref{eq:virial} with the last term omitted is
\begin{align}
r_{th,deg} = 2.4 \times 10^{-4} N_X^{1/6} (\rm{GeV}/m_X)^{1/2}~\rm{cm}. \label{eq:rthdeg}
\end{align}

\emph{Non-degenerate.}
If less than $N_{deg}$ particles have collected before collapse, the kinetic energy is simply $E_k^{non-deg} \sim 3 k_B T /2$ and the pre-collapse thermal radius is
\begin{align}
r_{th,nondeg} = 80 (\rm{GeV}/m_X)^{1/2} ~\rm{cm},
\end{align}
again assuming a neutron star temperature of $10^4~\rm{K}$.

\subsection{Critical DM number for collapse}

There are four states from which the collapse of DM can begin: the dark matter may be either degenerate or non-degenerate and either strongly-screened or Coulombic. We can determine which state precedes collapse as a function of model parameters ($\alpha_D,m_X,m_\phi$) using Eq.~\eqref{eq:virial}. 

\emph{Degenerate.} If more than $N_{deg}$ particles have collected, the dark matter will be degenerate. Substituting $E_k^{deg}$ into Eq.~\eqref{eq:virial}, and dropping the baryonic gravity term (which should be negligible after collapse begins), we can determine what parameter space initiates Coulombic collapse by requiring $y \leq 2$ in
\begin{align}
(3 \pi^2)^{2/3} m_\phi^2 / 2 m_X y^2 = 4 \pi \alpha_D m_\phi /y^3, \label{eq:degcoulvir}
\end{align}
from which we find that collapse will begin from a Coulombic (strongly-screened) state when $\alpha \gtrsim m_\phi/m_X$ ($\alpha \lesssim m_\phi/m_X$). If collapse begins from a Coulombic, degenerate state, the number of particles required for collapse can be found by solving Eq.~\ref{eq:virial} for $N_{coll}$,
\begin{align}
N_{coll}^{degCoul} > 10^{25} \alpha^{-6} (m_\phi/\rm{MeV})^{12} (\rm{GeV}/m_X)^9. \label{eq:ndegcoll}
\end{align}
If on the other hand, collapse begins from a strongly-screened degenerate state, Eq.~\eqref{eq:virial} must be solved numerically with $V_{Yuk}^{strong}$. Specifically, we solve Eq.~\eqref{eq:virial} for $y$, using successively larger values of $N_{coll}$, until no solution can be obtained where $y>2$, indicating collapse has begun. More details of this numerical solution can be found in \cite{Bramante:2013nma}.

\begin{figure}[h!]
\begin{center}
\includegraphics[scale=.9]{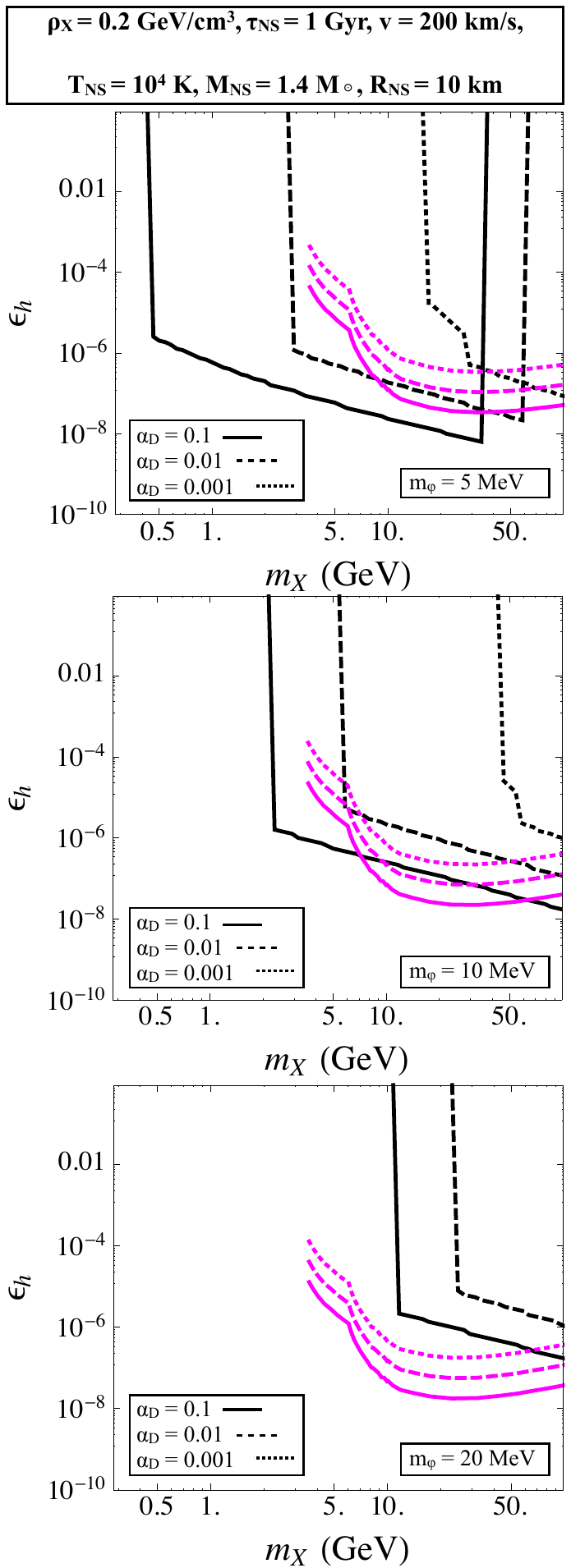}
\end{center}
\caption{Constraints from SuperCDMS and LUX (pink) along with bounds from old pulsars (black) on asymmetric Higgs portal dark matter are shown as a function of $\epsilon_h \equiv \sqrt{2}av/m_h^2$, for a number of values of $\alpha_D$ and $m_\phi$. The DM density ($\rho_X$), pulsar lifetime ($\tau_{NS}$), velocity dispersion ($v$), and pulsar temperature ($T_{NS}$), mass ($M_{NS}$), and radius ($R_{NS}$) used to plot the black lines are given at the top of the plot. Note that parameter space along each black line would collapse pulsars older than $\sim10^6$ years inside the Milky Way's central 500 parsecs, while permitting $\sim10^{9}$ year old pulsars outside of the galactic center.}
\label{fig:bounds2}
\end{figure}	

\emph{Non-degenerate.} If less than $N_{deg}$ particles have collected, dark matter will not be degenerate, and the kinetic energy will be $3k_B T/2$. For all parameter space of interest ($m_X \gtrsim 0.1 ~\rm{GeV}$), the number of particles required to begin collapse in a Coulombic, non-degenerate state exceeds $N_{deg}$ \cite{Bramante:2013nma}. Hence, we only need to consider collapse from a strongly-screened non-degenerate state. To determine the critical number of particles required, we insert $E_k^{non-deg}$ in Eq.~\eqref{eq:virial} and solve numerically for $N_{coll}$. As collapse proceeds, and $y$ (along with $r$) shrinks, the fermions will eventually become degenerate. The value of $y$ at which the fermions become degenerate can be found by substituting $r_{th,deg}$ into $y$,
\begin{align}
y_{deg} = 20 (m_\phi/\rm{MeV})(\rm{GeV}/m_X)^{1/2}. \label{eq:ydeg}
\end{align}
If $y_{deg} < 2$, we can substitute Eq.~\eqref{eq:ydeg} into Eq.~\eqref{eq:degcoulvir}, and find that collapse continues through a Coulombic, degenerate state so long as
\begin{align}
\alpha > 0.02(m_\phi/\rm{MeV})^2(\rm{GeV}/m_X)^{3/2}.
\end{align}
If $y_{deg} > 2$, then substituting Eq.~\eqref{eq:ydeg} into Eq.~\ref{eq:virial} (dropping the baryonic term), we find that collapse proceeds so long as
\begin{align}
\alpha > 10^{-6} e^{y_{deg}}(\rm{MeV}/m_\phi)(1+1/y_{deg})^{-1}.
\end{align}

After dark matter collapses into a black hole, in order to destroy the neutron star, it must accrete surrounding baryons faster than it radiates via the Hawking mechanism \cite{Hawking:1974sw}. This condition is fulfilled if,
\begin{align}
-1/(15360 \pi G^2 N_{coll}^2 m_X^2) + 4 \pi \rho_B G^2 N_{coll}^2 m_X^2/c_s^3 > 0,
\label{eq:accret}
\end{align}
where the first term is the Hawking radiation rate, the second term is the Bondi accretion rate, and $c_s \sim 0.3$ is the sound speed of baryons in the neutron star. If the black hole grows, it will destroy the neutron star in $t_{des} \sim 2000 (\rm{GeV}/m_X)(10^{40}/N_{coll})~\rm{yrs}$.

In Figure \ref{fig:epsvsmx} and \ref{fig:bounds2}, we display parameter space which is excluded by the existence of $10^9$ year old pulsars outside the galactic center (where the DM density is $\rho_{X} \sim 0.2 ~\rm{GeV/cm^3}$). Note that the left side of these bounds terminate where $N_{coll.} > C_X \tau_{NS}$, while the right side terminates when the condition Eq.~\eqref{eq:accret} fails, meaning the black holes formed by collapsing dark matter are too small too grow and instead evaporate. In Figure \ref{fig:epsvsmx}, we show model space which would collapse $10^6-10^7$ year old pulsars in the galactic center, while allowing for old pulsars outside the galactic center. Note that we assume a galactic center DM density ($\rho_X \sim 80 ~\rm{GeV/cm^3}$) determined by DM self-interactions, as we discuss in the next section.

\section{Dark matter self-interactions and pulsar age curves}
\label{sec:agecurves}

To determine which Higgs portal PCDM models precipitate pulsar destruction in the galactic center, we need to know the mass distribution of DM in the Milky Way. It is commonly assumed that DM is cold and collisionless (CCDM), and simulations of CCDM structure formation yield halo densities that follow the Navarro-Frenk-White (NFW) profile: $ \rho (r) = \rho_0 /(r/R_s) \left(1+ r/R_s\right)^{-2} $, where $\rho_0$ and $R_s$ are characteristic density and scale radius respectively, and $r$ is the distance from the galactic center. However, the CCDM paradigm has been challenged recently by some observations of small scale structure. These include dwarf halo profiles which appear underdense at their centers \cite{Oh:2010ea,KuziodeNaray:2007qi}, too few small satellite galaxies near the Milky Way, and no large satellite galaxies \cite{Sawala:2010zw,BoylanKolchin:2011de,BoylanKolchin:2011dk}. 

These small scale structure anomalies can be addressed with the addition of sizable DM self-interactions $\sigma_{SI} \sim 10^{-24} ~\rm{cm^2} / \rm{GeV}$. However, this seems disfavored by self-interaction bounds from ``bullet" collisions in galactic clusters \cite{Randall:2007ph,Harvey:2015hha} and the observed ellipticity of galactic cluster cores \cite{MiraldaEscude:2000qt}. Typically these require DM self-interactions smaller than $\sigma_{SI}/m_X \sim 10^{-24} ~\rm{cm^2} / \rm{GeV}$. If, on the other hand, DM self-interactions are velocity dependent through coupling to a light mediator ($m_\phi=1-100~\rm{MeV}$), bounds from large scale observations can be evaded \cite{Tulin:2012wi,Tulin:2013teo,Hooper:2008im,ArkaniHamed:2008qn,Pospelov:2008jd,Kaplan:2009ag,Buckley:2009in,Loeb:2010gj}. This is because the cross-section can be large in, for example dwarf galaxies, where the characteristic velocity is $50 ~\rm{km/s}$, while the mediator is more off-shell and interactions are weaker for DM velocities $\sim 1000 ~\rm{km/s}$, in galactic clusters. Much of the parameter space shaded grey in Figure \ref{fig:epsvsmx} will have velocity-dependent self-interactions suitable for resolving small scale anomalies, while remaining consistent with large scale bounds \cite{Tulin:2013teo,Kaplinghat:2013yxa}.

\begin{figure}[h!]
\begin{center}
\includegraphics[scale=.8]{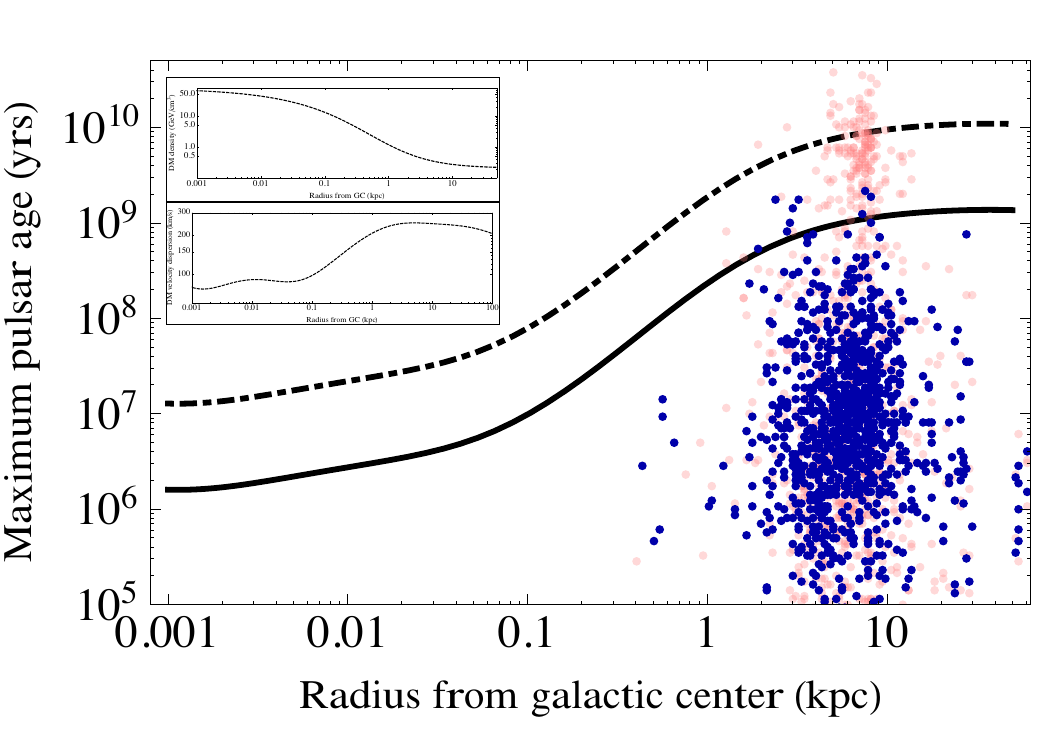}
\end{center}
\caption{The maximum age of an $R_{NS} = 10 ~\rm{km}$, $M_{NS} = 1.4 M_\odot$, $T_{NS} = 10^4 ~ \rm{K}$ pulsar is plotted as a function of distance from the galactic center. The solid and dotted-dashed maximum age curves result from parameter space shown in Figure \ref{fig:epsvsmx}, also indicated in Figure \ref{fig:epsvsmx} plots with solid and dotted-dashed curves. For example, the solid curve is consistent with light dark matter ($m_X = 6 ~\rm{GeV}$) strongly coupled ($\alpha_D = 0.1$) to a light scalar ($m_\phi = 15 ~\rm{MeV}$) with a small mixing ($\epsilon_h = 10^{-6}$) with the Higgs boson; this parameter space lies along the solid black curve in the top panel of Figure \ref{fig:epsvsmx}. Pulsar data points are overlaid; ``characteristic" ages ($\tau_p \lesssim P/2\dot{P}$) and radial distances were taken from the ATNF pulsar catalog \cite{Manchester:2004bp}. Note that characteristic ages should be treated as upper bounds on pulsar lifetimes -- actual pulsar ages can be significantly shorter, especially if $P_0 \sim P$ where $P_0$ is the initial pulse period. (Indeed, many pulsar characteristic ages above the dotted-dashed curve exceed the age of the universe.) Pulsar data points shown in dark blue were required to have longer pulse periods $P \gtrsim 0.5 ~\rm{s}$ and to be solitary (not in a binary); these conditions may indicate better alignment between characteristic and actual age \cite{Tauris:2015bra}. Inset are the dark matter density (top) and velocity dispersion (bottom) as a function of distance from the galactic center.}
\label{fig:agecurves}
\end{figure}

Some simulations indicate that DM with significant self-interactions $\sigma_{SI} \gtrsim 10^{-24} ~\rm{cm^2} / \rm{GeV}$, will cause the Milky Way's DM density profile to be flat ($\sim \rm{GeV/cm^3}$) within the central 2 kiloparsecs \cite{Vogelsberger:2012ku,Peter:2012jh,Rocha:2012jg}. However, a recent study which included the gravitational potential of baryons showed that SIDM in Milky Way size halos will create a 500 parsec DM core, with a larger constant central density ($\sim 80 ~\rm{GeV/cm^3}$) \cite{Kaplinghat:2013xca}. For SIDM, the density of the central core will depend mostly on central the baryonic density, $\rho (r) = \rho_0 \exp[ h( r)]$, where $h( r)$ satisfies Jean's equation and is given by
\begin{align}
h(r) = \frac{- 2 \pi G \rho_0 r_0^2}{\sigma_0^2}y\left(\sigma_0-\frac{2y}{3}\right). \label{eq:dmprofile}
\end{align}
In the Milky Way, $\rho_0 = 80 ~\rm{GeV/cm^3}$ is the central DM density, $ y = r/(r+r_0)$ parameterizes distance from the galactic center, $r_0 =  2.7 ~\rm{kpc}$, $\sigma_0 = 160 ~\rm{km/s}$ is a constant velocity dispersion. We show this DM density profile in the upper left inset of Figure \ref{fig:agecurves}, along with the DM velocity dispersion we assumed, which was derived by fitting observed star velocities in the Milky Way \cite{Sofue:2013kja, Bramante:2014zca}.

In Figure \ref{fig:agecurves} we use the DM density profile in Eq.~\eqref{eq:dmprofile} to predict the maximum pulsar age as a function of distance from the galactic center. The solid (dotted-dashed) line shows the maximum age curve for asymmetric Higgs portal models that predict $\gtrsim 10^6$ ($\gtrsim 10^7$) year old pulsars will collapse in the galactic center. The radial distance from the galactic center and the characteristic radii of pulsars in the Milky Way, shown with blue and pink data points, are taken from the ATNF pulsar catalogue \cite{Manchester:2004bp}. It should be pointed out that the characteristic age of pulsars $\tau_{p} \lesssim  P/2\dot{P}$, often vastly overestimates the age of the pulsar. Indeed, it has become clear that this overestimate may be systematically large for all millisecond pulsars with $\tau_p \gtrsim 10^9$ years \cite{Tauris:2012ex,Tauris:2015bra}. This effect results from an initial pulse period, $P_0$, that is not much smaller than the current pulse period $P$. In this case the full expression for the pulsar age, $\tau_{p}^{full} \simeq  P/2\dot{P} - P_0/2\dot{P}$, is required, but $P_0$ is unknown. Note in Figure \ref{fig:agecurves} that many of the pulsar ``characteristic" ages exceed the age of the universe. 

However, millisecond pulsars that formed in a binary system with a white dwarf (WD) provide an alternative way to check the age of the pulsar. The binary white dwarf's temperature and mass can be fit to WD cooling curves to independently determine the age of the binary system. One example is pulsar J1738$+$0333, which has a characteristic age of 4 Gyr, and whose white dwarf has an apparent age of $0.5-5$ Gyr \cite{Antoniadis:2012vy}. Therefore, for the purpose of bounding DM parameters with pulsar ages, it is most conservative to assume pulsars outside the galactic center have reached ages of about a gigayear.

\section{Conclusions}
\label{sec:conclusions}

Higgs portal DM-induced pulsar collapse could explain the apparent paucity of galactic center pulsars. Fermionic asymmetric dark matter coupled to the SM through a Higgs portal can collapse pulsars older than $\sim 10^6$ years inside the galactic center, while not collapsing $10^9$ year old pulsars near the solar position. This is because the expected DM density in the galactic center ($\rho_{X} = 80 ~\rm{GeV/cm^3}$) is larger than in the rest of the halo ($\rho_{X} = 0.2 ~\rm{GeV/cm^3}$), leading to increased DM capture in GC pulsars.

Based on the number of GC pulsar progenitor stars, GC radio surveys should have already found $O(10)$ pulsars in the central parsec. However, none have been observed, and it is unlikely that this is solely the result of measurement limitations \cite{Dexter:2013xga}, although it could result from an overestimated GC pulsar population. We have showed that for DM masses $m_X = 0.1-100 \rm{GeV}$, with Higgs portal mediator masses $5-20 \rm{MeV}$, DM-mediator couplings $\alpha_D = 0.001-0.1$, and mediator-Higgs mixings $\epsilon_h = 10^{-8} -10^{-2}$ asymmetric Higgs portal dark matter provides an explanation for the absence of GC pulsars: older GC pulsars collapse into black holes after enough DM is captured to form a black hole that will grow in pulsar interiors. As pulsars are discovered in the galactic center by future radio surveys and radio telescopes, (e.g. FAST \cite{Nan:2011um} and the Square Kilometer Array \cite{Eatough:2015jka}) PCDM would manifest as a maximum pulsar age that increases with distance from the GC. 

Applied to old pulsars seen outside the GC, our results set bounds on fermionic, asymmetric Higgs portal dark matter that are often more stringent than those set by direct detection experiments. However, while much of the Higgs portal PCDM parameter space is inaccessible to present terrestrial direct detection, we have shown that next-generation direct detection experiments will either find or exclude it. In addition, through its coupling to a light mediator, Higgs portal PCDM naturally fits one explanation for the core-cusp phenomenon: light mediator DM self-scattering is resonantly enhanced at smaller momenta in dwarf galaxies, and is diminished at larger momenta in spiral galaxies and galactic clusters.

Finally, we note that a recent proposal of Fuller and Ott \cite{Fuller:2014rza} (advanced with all reserve) has linked PCDM to fast radio bursts \cite{Lorimer:2007qn,Keane:2012yh,Thornton:2013iua,Petroff:2014taa,Spitler:2014fla}. We leave this and other complementary probes of PCDM to future work.

\acknowledgements
We have benefited from discussions with Asimina Arvanitaki, Gordan Krnjaic, Adam Martin, Tim Linden, James Unwin, James Wells, and Itay Yavin. JB is grateful to the CERN theory division for hospitality while portions of this work were completed. This research was supported in part by Perimeter Institute for Theoretical Physics. Research at Perimeter Institute is supported by the Government of Canada through Industry Canada and by the Province of Ontario through the Ministry of Economic Development \& Innovation.

\appendix
\section{Notes on the galactic center pulsar population}
\label{app:pulsars}

This appendix details methods for estimating the number of pulsars, both young and millisecond, at the galactic center. 

\emph{Young pulsars in the central parsec.} A simple way to estimate the expected number of young pulsars in a region is to count the number of 8-20 solar mass, high mass stars, $N_{hms}$. It is expected that the majority of these will expire in core collapse supernovae and form pulsars \cite{Pfahl:2003tf,Dexter:2013xga}. More massive stars have larger hydrogen cores which burn faster, and so as a general rule the luminosity of stars scales as $L \sim M^{3.5}$. Normalizing to a sol-type star, the lifetime of a high mass star ($t\sim M/L$) is given by $t_{hms} = 10^{10} (M/M_{\odot})^{-2.5} ~\rm{yrs}$. Assuming that the abundance of each star type stays constant in the region of interest, this implies there are $\sim 5N_{hms}/3 $ young pulsars, because the lifetime of the high mass stars is $6$ Myr, and a typical lifetime for a young pulsar is $10$ Myr. 

It follows that the inner parsec of the galactic center which hosts $\sim$300 high mass stars should also host $\sim$500 young pulsars, $\sim$50 of which would beam towards earth, $\sim$25 of which should have already been detected given the recently measured radio scattering dispersion (courtesy of a bright magnetar) \cite{Dexter:2013xga}. However, this calculation is only valid if high mass stars in the galactic center were as abundant historically as they are now; high mass stars may have been less abundant 20 million years ago \cite{Bartko:2009qn}. The next generation of surveys should permit imaging of 1-2 solar mass stars \cite{Schodel:2014wma}, which as the byproducts of high mass stars, will provide a better indication of the historical abundance of high mass progenitor stars.

\emph{Older, millisecond pulsars.} Millisecond or recycled pulsars form when a neutron star spins up in a binary system by accreting gas from its companion \cite{Tauris:2015bra}. Most millisecond pulsars have been found inside globular clusters, which have stellar densities as large as $1-10^3$ solar masses per cubic parsec, comparable to a stellar density of $10^3 M_\odot/\rm{pc^3}$ and $10^6 M_\odot/ \rm{pc}^3$ in the central 100 parsecs and central parsec of the milky way, respectively. One way to estimate the central parsec millisecond pulsar population is to scale up the millisecond population in globular clusters (about 10-50 millisecond pulsars per cluster) \cite{Gordon:2013vta}. Millisecond pulsar formation is expected to be greater in the central parsec, because of its higher stellar density and higher escape velocity (400 vs. 50 km/s), allowing it to retain a larger fraction of young neutron stars (with an average birth velocity of 400 km/s). Given these factors, $\sim$1000 milliscond pulsars are expected in the central parsec \cite{FaucherGiguere:2010bq,Abazajian:2012pn,Gordon:2013vta}.

We can also consider a simpler, more conservative method for estimating the millisecond pulsar population of the central parsec. Before a neutron star has finished accreting gas from its companion, it will necessarily be in a low mass x-ray binary. As gas falls into the compact neutron star, it often emits x-rays. Hence, we should expect the observation of low mass x-ray binaries to correlate with the number of millisecond pulsars. Indeed, studies have shown that both of these correlate with the stellar encounter rate in globular clusters, $\Gamma_c \sim \rho_c r_c^3 v_c^{-1}$, where these variables are respectively the density, radius, and linear velocity dispersion of a globular cluster \cite{Bahramian:2013ihw,Verbunt:2013kka}. Roughly $10-20$ times more millisecond pulsars than low mass x-ray binaries have been found in globular clustars \cite{Bahramian:2013ihw}. Extrapolating from the four x-ray binaries found in the central parsec \cite{Muno:2004xi,Degenaar:2012tm}, one would expect $\sim$40-80 detectable millisecond pulsars in the central parsec, of which (accounting for increased radio pulse dispersion at the galactic center) $\sim$5-10 would have been detected by the 14 Ghz survey \cite{Macquart:2010vf,Dexter:2013xga}.

\section{$\sigma_{nX}$ velocity dependence for Higgs portal DM}
\label{app:snx}

The matrix element for dark matter t-channel scattering off nuclei is
\begin{equation}
i {\cal M} = g_D \epsilon_{N}  \frac{[\bar u(p_4) {u}(p_1)][\bar u(p_3) u (p_2)]}{t-m_\phi^2+i\epsilon},
\end{equation}
where $p_{1,3}$ and $p_{2,4}$ are the dark matter and nucleon initial and final 4-momenta, respectively. The prefactor $\epsilon_{N} \simeq 3 \times 10^{-3} \times \epsilon_h$ parameterizes the coupling of $\phi$ to nuclei, largely determined by the gluon hadronic matrix element \cite{Kanemura:2010sh}.

In the center momentum frame of the interaction, the four momenta are given by 
\begin{align}
p_{1} &= (E_{1},0,0,p_{cm}) \nonumber
\\
p_{2} &= (E_{2},0,0,-p_{cm}) \nonumber
\\
p_{3} &= (E_{3},-p_{cm}{\rm sin~\theta},0,-p_{cm}{\rm cos~\theta}) \nonumber
\\
p_{4} &= (E_{4},p_{cm}{\rm sin~\theta},0,p_{cm}{\rm cos~\theta}).
\end{align}
To approximate the cross-section at a direct detection experiment or in a neutron star, we integrate the square of the amplitude,
\begin{align}
&\frac{1}{4}|\mathcal{M}|^2 =  \nonumber 
\\
& g_D^2  ~\epsilon_{N} ^2  ~\frac{4 (2 m_x^2 +  p_{cm}^2(1- \cos ~\theta))(2m_n^2 + p_{cm}^2(1- \cos ~\theta))}{(-2 p_{cm}^2(1- \cos ~\theta) - m_\phi^2)^2},
\end{align}
over a range of nuclear recoil energies $E_{\rm recoil}$, specifically ranging over incoming DM momenta $0.001 \gamma v m_X$ to $\gamma v m_X$, where $\gamma = 1/\sqrt{1- v^2}$, and $v$ is a typical DM velocity in the nucleon's rest frame ($v \simeq 10^{-3}~ \rm c$ and $0.7 ~\rm c$ for DD experiments and pulsars, respectively). Here we define the momentum transfer $ Q \equiv \sqrt{p_{cm}^2 (1-\cos \theta)}$, which can be related to the recoil energy with $ p_{cm}^2 (1-\cos \theta) \sim 2 m_n E_{\rm recoil}$. 

Some recent work has shown that Dirac dark matter with a strong ($\alpha_D \gtrsim 0.1$) coupling to a light scalar mediator, may form bound states at freeze-out \cite{Wise:2014jva,Wise:2014ola}. This bound state dark matter can have a cross-section enhanced by $N_D^2$, the square of the number of bound fermions, as long as the momentum transfer of the interaction is less than the binding energy $Q \ll BE_0$. For the purposes of this study, we assume no bound state enhancement to the cross-section -- this assumption is certainly valid for the capture of DM in neutron stars where $Q \sim m_X \gg BE_0$. However, there may be some enhancement to direct detection capture when $\alpha_D^2 \gg 4 m_\phi / m_X$, which would shift direct detection bounds on $\epsilon_h$ in Figures \ref{fig:epsvsmx} and \ref{fig:bounds2}.

\bibliographystyle{JHEP.bst}

\bibliography{portals}

\end{document}